\documentclass[10pt, twocolumn]{article}
 
\title{\textbf{Access Timing as Scaffolding: A Reinforcement Learning Approach to GenAI in Education}}

\author{
    \fontsize{11}{13}\selectfont 
    1\ts{st} Janne Rotter \\
    \fontsize{10}{11}\selectfont 
    Universitat Pompeu Fabra\\
    \fontsize{10}{11}\selectfont 
    Barcelona, Spain\\
    \fontsize{10}{11}\selectfont   jannedavid.rotter01@estudiant.upf.edu\\
    \fontsize{10}{11}\selectfont 
    ORCID: 
    \href{https://orcid.org/0009-0001-3520-958X}{ 0009-0001-3520-958X}
    \and
    \fontsize{11}{13}\selectfont 
    2\ts{nd} Pau Benazet i Montobbio \\
    \fontsize{10}{11}\selectfont 
    Universitat Pompeu Fabra\\
    \fontsize{10}{11}\selectfont 
    Barcelona, Spain\\
    \fontsize{10}{11}\selectfont   pau.benazeti01@estudiant.upf.edu\\
    \fontsize{10}{11}\selectfont 
    ORCID: 
    \href{https://orcid.org/0009-0007-7759-6016}{0009-0007-7759-6016}
    \and
    \fontsize{11}{13}\selectfont 
    3\ts{rd} Davinia Hern\'{a}ndez-Leo \\
    \fontsize{10}{11}\selectfont 
    Universitat Pompeu Fabra\\
    \fontsize{10}{11}\selectfont 
    Barcelona, Spain\\
    \fontsize{10}{11}\selectfont   davinia.hernandez-leo@upf.edu\\
    \fontsize{10}{11}\selectfont 
    ORCID: 
    \href{https://orcid.org/0000-0003-0548-7455}{0000-0003-0548-7455}
}

\date{}

\usepackage{lipsum}
\usepackage{lmodern} 
\usepackage{array} 
\usepackage{booktabs} 
\usepackage{dblfloatfix}

\usepackage[
layout=letterpaper, 
paper=letterpaper, 
portrait, 
head=0.5in,
foot=0.5in,
top=1in, 
bottom=1in, 
left=0.75in, 
right=0.75in
]{geometry}

\usepackage{tabularx}
\usepackage{booktabs} 
\usepackage{longtable} 
\usepackage{caption}
\usepackage{subcaption}

\usepackage[english]{babel}
\renewcommand{\thesection}{\arabic{section}}
\renewcommand{\thesubsection}{\thesection.\arabic{subsection}}

\usepackage{enumitem}
\setlist{itemsep=0.3em, topsep=0.3em}
\usepackage{xurl}

\usepackage[runin]{abstract}

\usepackage{hyperref}

\hypersetup{
colorlinks=true,
urlcolor=blue,
linkcolor=blue,
citecolor=blue,
pdftitle={Access Timing as Scaffolding: A Reinforcement Learning Approach to GenAI in Education - Rotter et al.},
pdfsubject={Original Article},
pdfauthor={Rotter et al.},
pdfkeywords={Generative AI in Education, Reinforcement Learning, Metacognition, Intelligent Tutoring Systems, Productive Failure, AI Access Timing}
}

\usepackage[backend=biber,style=numeric,sorting=none]{biblatex}
\usepackage{etoolbox}
\usepackage{indentfirst}

\usepackage{custom_commands} 

\usepackage{subfiles} 

\usepackage[pdftex]{graphicx} 

\begin{document}
\renewcommand{\abstractname}{}

\maketitle 


\setlength{\absleftindent}{0em}
\setlength{\absrightindent}{0em}

\begin{abstract}
    \abstractText 
\end{abstract}

\section{Introduction}
\label{chap:intro}
Generative Artificial Intelligence (GenAI) has revolutionized the way we learn and understand within higher education contexts \cite{hernandez2025collaborating}. While historical innovations of educational technologies such as films and interactive whiteboards have already reshaped the way education was done in the past, GenAI represents a qualitatively different tool that can autonomously generate ideas, texts, and solutions \cite{noy2023experimental}. Modern GenAI applications offer support from source finding to linguistic polishing and are omnipresent in most students’ daily lives, regardless of institutional policies \cite{yan2024promises}. According to a survey by the digital education council amongst 3,800 university students across 16 countries, already 86\% of them used AI for their studies in 2024 \cite{dec2024ai}.

While these systems offer a number of upsides such as individualized experience and unprecedented access to information, they also have a tendency to increase students’ overreliance and reduce their critical thinking capacities \cite{gerlich2025ai}. University students themselves notice this trend, with over 50\% believing that over-reliance on AI negatively impacts their academic performance \cite{dec2024ai}. Similarly, an OECD study reports that while AI use can improve performance on practice tasks, students may perform worse on exams where AI is unavailable, suggesting that apparent learning gains may mask growing over-reliance \cite{oecd2026outlook}. Paradoxically, completely banning GenAI use may not effectively address this issue and could even have unintended consequences, such as increasing unreflective use \cite{lim2023generative}. Therefore, there is a growing degree of agreement that the aim should not be to completely prohibit GenAI use, but rather to explore how it can be applied in ways that position AI and learners as collaborators, with their interactions informed by pedagogical and ethical frameworks \cite{lim2023generative, hernandez2025collaborating}. The primary question therefore becomes not how to best prohibit GenAI use for learning, but when to allow it to foster pedagogical goals. 

\subsection*{The Present Study}

While most research focuses on how to explicitly scaffold generative AI use, we investigate a complementary dimension that received far less attention: when access to GenAI should be provided and how it affects learning outcomes. Recent work by \cite{zhi2026investigating} has begun to treat the timing of LLM access as an empirical variable in its own right and compared fixed early, continuous, late, and no-LLM access conditions on a critical thinking task. While they found that the effect of access timing depended on available task time, their approach was fixed rather than adaptive, focused on in-session task performance and was not grounded in pedagogical theory, ultimately leaving the question open whether theoretically motivated, adaptive timing can support durable learning. 

Reinforcement learning offers a promising framework to operationalize the question of when pedagogically desirable access should be granted, as it enables an agent to learn when to adaptively intervene to maximize educational outcomes and balance different learning objectives \cite{fahad2023reinforcement}. Furthermore, its clear concept of agent actions and states is easily applicable as an online episodic task description to the question of when to intervene.

Therefore, we develop a prototype that implements pedagogically grounded access timing and enables its empirical evaluation. We use reinforcement learning as a mechanism to derive an adaptive access policy informed by educational theory. Overall, we aim to answer the following research questions:

\begin{itemize}
\item \textbf{RQ1}: Can a pedagogically grounded policy for when to allow GenAI improve learning gains (post-test performance) compared to naive baselines (always allow/never allow)?
\item \textbf{RQ2}: How does adaptive timing of AI access influence students’ metacognitive skills?
\end{itemize}

In line with contemporary research focused on pedagogical steering \cite{puech2024towards, fahad2023reinforcement}, the reinforcement learning agent utilizes pedagogical reward shaping, student modelling, and Proximal Policy Optimization to learn when to allow the usage of generative AI in the learning process to foster knowledge acquisition and metacognitive engagement. 
The agent thus serves as a theoretically grounded mechanism to operationalize pedagogical frameworks into a single, adaptive access policy, which enables an empirical test of whether timing of GenAI access improves learning outcomes. To evaluate the developed solution we focus on text comprehension, as it requires learners to actively construct meaning and monitor their understanding, making it well suited for studying how the timing of AI assistance influences both learning outcomes and metacognitive engagement.

\section{Theoretical Background}
\label{chap:theory}

\subsection{Generative AI in Education}

GenAI can be used for a wide range of use cases in the educational domain, such as grading assistants \cite{jauhiainen2025generative}, delivering individualized feedback \cite{becerra2024generative}, or generating new content like quizzes \cite{li2024generating} and educational videos \cite{thareja2024eden}. Furthermore, GenAI can be used to facilitate personalized on-demand support via hints \cite{patikorn2020effectiveness}. However, while a number of studies attest a positive effect of on-demand support on educational outcomes \cite{stamper2013experimental, ostrow2014testing}, a meta-analysis by Kluger \& DeNisi \cite{kluger1996effects} indicates that providing too detailed feedback can result in diminished learning gains, which suggests that only selectively allowing hints can be beneficial. This highlights the tension between pedagogical opportunities and risks. 

Generally, there is a consensus in academic thought both being aware of the potential upsides of GenAI in education and acknowledging its prospective pitfalls. For instance, across the literature, a common thread is the recognition that GenAI can enhance education by enabling personalization, fostering interactivity, delivering timely feedback, and supporting innovative teaching and assessment practices \cite{yan2024promises, baidoo2023education, mittal2024comprehensive, giannakos2025promise}. On the other hand, scholars consistently highlight the ethical, technical, and pedagogical challenges posed by GenAI, emphasizing concerns over accuracy, bias, privacy, accountability, and the disruption of established educational practices \cite{yan2024promises, baidoo2023education, mittal2024comprehensive}.

Given this duality of promise and risk, the central pedagogical question has shifted from whether to integrate GenAI into education toward understanding when and how its use can most effectively support learning.

\subsection{Pedagogical Background}
\noindent
In order to be able to inform when to allow GenAI use to facilitate learning, reduce over-reliance and foster metacognitive growth a thorough understanding of the pedagogical literature body is required. Central to this understanding is the concept of scaffolding which describes the provision of a support mechanism that enables a learner to solve a problem or carry out a task which would otherwise be beyond their unassisted effort \cite{wood1976role}. In educational technology, scaffolding has been implemented through a wide range of mechanisms such as adaptive hints \cite{phung2025plan} or structured feedback \cite{feng2024application} . Most prior work on scaffolding in AI-supported learning focuses on \textit{how} support is delivered. By contrast, this work investigates the complementary dimension of \textit{when} access to GenAI support is granted, treating access timing itself as a form of implicit scaffolding.

\noindent
\subsubsection*{Metacognition and Generative AI}

Metacognition is generally defined as "thinking about your own thinking" \cite{flavell1979metacognition, jaleel2016study, mahdavi2014overview}. It refers to our understanding of how we plan and conduct our learning and how cooperation with tools influence this process. For instance, after writing an exam, a student might realize that rereading and learning in a short time (cramming) were not the most effective study strategies to prepare. Reflecting on this experience, they may decide to adjust their study strategies for future exams. For instance, they might decide to alternate between different learning strategies, a process known as interleaving \cite{ostrow2015blocking, rohrer2020randomized}. This is a classic example of metacognition in action \cite{rivers2020measuring}. Students’ metacognitive abilities are of significant relevance in educational contexts, as they are a meaningful predictor for their learning and performance \cite{stanton2021fostering}, even outweighing factors like intelligence \cite{veenman2005relation}. Furthermore, building metacognitive skills in learners often leads to an increase in self-confidence \cite{jaleel2016study}. It is thus often suggested that metacognitive abilities distinguish good from bad learners \cite{jaleel2016study, mahdavi2014overview}.

GenAI, however, is a disrupting factor that can influence metacognitive elements, especially regarding younger students who did not have sufficient time to form a metacognitive understanding of their learning process \cite{gerlich2025ai}. For instance, Fan et al.  found that GenAI  encourages learners to rely heavily on technology and may lead to metacognitive “laziness” \cite{fan2025beware},  which could undermine their metacognitive control and the capacity to self-regulate and engage in critical thinking. Furthermore, the instant response common to GenAI models and insufficient attention to developing metacognitive capabilities and reflective practices \cite{hou2025role} may also lead to the development of a dependency in students, who consequently no longer actively learn but passively accept GenAI outputs, reducing their critical thinking ability and metacognitive engagement \cite{chan2023comprehensive, yang2025analysing, zhai2024effects}. Similarly, it was shown that using AI can significantly hinder the metacognitive ability to accurately assess ones own performance, with high AI literacy being a primary predictor of low accuracy \cite{fernandes2025ai}. As a result, students' self-regulated learning abilities in GenAI environments can be enhanced by external metacognitive support \cite{xu2025enhancing} or integrating metacognitive support strategies \cite{tankelevitch2024metacognitive}, whereas an absence of such strategies has negative effects on learning outcomes \cite{xu2025enhancing}.

In line with these findings, creating Intelligent Tutoring Systems (ITS) that facilitate metacognitive reflection is an ongoing research area. ITS are particularly well fit for metacognitive focused education as they can tailor content to individual students. This is important as different learning styles play a crucial role in how students engage with content \cite{azevedo2022lessons}. Consequently, adapting metacognitive strategies to students via ITS can facilitate effective learning \cite{azevedo2022lessons, cantero2025instructional}. While some studies did not find a significant impact of increasing metacognitive engagement in ITS on educational outcomes \cite{mccarthy2018metacognitive}, most research suggests that metacognitive prompts can significantly enhance learning  \cite{abdelshiheed2023leveraging} and the ability to seek help when needed \cite{roll2011improving}.

A commonly used framework for metacognition in Schraw \& Dennison's \cite{schraw1994assessing} differentiation between metacognitive knowledge, the awareness of one’s thinking, and metacognitive regulation, the ability to manage one’s own thinking processes. To investigate metacognition rigorously, this framework will be used in the context of this work. 

\noindent
\subsubsection*{Cognitive Offloading} 
Cognitive Load Theory (CLT) provides another important perspective for examining the role of GenAI in educational contexts \cite{sweller1988cognitive}. It proposes that the human cognitive system has a finite capacity, and lowering cognitive load can improve both learning and performance while also reducing mental strain \cite{gerlich2025ai}. One way to achieve this is through cognitive offloading, which involves using external tools to lessen working memory demands which can consequently lead to greater learning gains \cite{risko2016cognitive}. For instance, Zhao et al. \cite{zhao2025generative} showed that the introduction of GenAI can help students reduce the external cognitive load to some extent, freeing up more space for the processing, construction, and automation of schemas. Furthermore, the ability to remember planned future actions, known as prospective memory, can improve when individuals transfer or externalize their intentions \cite{gilbert2015strategic}. This process of cognitive offloading is also strongly associated with metacognitive decision-making \cite{gilbert2015strategic}.

However, there are also potential downsides to cognitive offloading. While it can free up cognitive resources, there is concern that it may lead to a reduction in internal cognitive abilities, such as memory retention and critical analysis skills \cite{gerlich2025ai}. For instance, drivers who use GPS initially navigate routes more efficiently than those without it, but they tend to drive more slowly on later attempts without GPS guidance \cite{fenech2010effects}. This effect occurs because GPS use can disrupt the acquisition of spatial knowledge by diverting attention \cite{gardony2015navigational} and showcases the dual nature of cognitive offloading. In the case of AI tools and search engines, studies have shown that the mechanism of cognitive offloading can potentially diminish memory retention and deep information processing \cite{sparrow2011google, firth2019online}.  

An issue that emerges as a significant concern regarding cognitive offloading is that its associated costs may not be immediately apparent to those affected \cite{atchley2024human}. 
In GenAI environments, students may place unjustified trust in these technologies and the consequently acquired skills \cite{zhai2024effects, klingbeil2024trust}. This phenomenon of overtrust was also shown to lead to greater cognitive offloading, ultimately resulting in diminished critical thinking \cite{klingbeil2024trust}. Taken together, cognitive offloading presents both benefits and drawbacks for learning, underscoring the importance of a balanced approach to using GenAI in education, leveraging the tool’s advantages while remaining mindful of the risks of over-reliance. 

\noindent
\subsubsection*{Productive Failure} 
Another crucial theoretical lens through which to view GenAI use in education is the productive failure framework. It emphasizes that permitting learners to engage in unstructured problem-solving and to experience struggle or failure before receiving elaborate explanation and scaffolding can positively influence learning outcomes \cite{kapur2008productive}. It further emphasizes that we tend to learn more effectively from our own failed attempts than from others’ \cite{kapur2014productive}. This framework differentiates between different types of learning: Productive failure refers to situations in which learners initially fail to solve a problem but gain valuable insights that enhance learning in the long run. In contrast, unproductive success describes cases where learners reach the right answer but without meaningful comprehension or transferable knowledge \cite{kapur2016examining}. Illustrating the practical importance of productive failure, a meta-study by Darabi et al. \cite{darabi2018learning} could attest to moderately positive result for the effect of learning from failure.

A number of cognitive explanations exist in the literature for this phenomenon. For instance, DeCaro \& Rittle-Johnson \cite{decaro2012exploring} and Schwartz et al. \cite{schwartz2011practicing} argue that engagement in productive struggle may help to activate and differentiate relevant prior knowledge. In relation to CLT, Kapur \cite{kapur2014productive} suggests that learning through productive failure involves a balance between the negative impact of increased cognitive load and the positive effects of activating and differentiating prior knowledge, with the benefits outweighing the drawbacks as long as the cognitive load does not become overwhelming enough to cause learners to disengage.

GenAI might take away this initial productive struggle as it positions information and correct solutions directly at our fingertips. For instance, a student might directly query a generative AI model before even trying himself, resulting in unproductive success \cite{pak2024productive}. This is supported by contemporary research. For instance, a mixed-method study amongst 381 students could show that a subset of students superficially copied AI-outputs, being an example of unproductive success that limits students’ ability to engage in critical thinking and independent problem-solving \cite{zhai2025evaluating}. This is particularly worrisome as many students do not possess the metacognitive awareness required to differentiate between effective and ineffective uses of AI support \cite{fan2025beware}. Regarding the challenge of the over-reliance of students on AI for problem-solving, decision-making, and cognitive tasks \cite{malik2023exploring, chan2024will} research also suggests that fostering a problem-based learning environment where students are encouraged to first use analytical reasoning before turning to AI solutions can be beneficial \cite{walter2024embracing}, which directly aligns with the notion of productive failure. Importantly, no studies to date have directly investigated the relationship between productive failure and GenAI, highlighting a significant gap in the literature. While not directly investigating this gap, this study utilizes productive failure to inform pedagogically grounded GenAI use. 

Taken together, these pedagogical perspectives suggest that while GenAI can reduce cognitive load and support learning, premature access may encourage cognitive offloading and reduce metacognitive engagement. Conversely, allowing learners to initially struggle with a task can promote deeper understanding through productive failure. These findings suggest that the access timing to GenAI assistance may be a crucial factor determining whether generative AI supports or undermines learning.

\subsection{Reinforcement Learning for Hint Giving in Education} 
Reinforcement learning in educational contexts is an active and growing research area \cite{riedmann2025reinforcement}.Generally, RL is a machine learning paradigm in which an agent learns to make decisions through interaction with an environment \cite{sutton2018reinforcement}. At each step, the agent observes the current state of the environment and selects an action according to a policy, which defines the agent’s strategy for choosing actions. After performing an action, the agent receives a reward signal that reflects the desirability of the outcome. Over time, the agent updates its policy to maximize the cumulative reward, allowing it to learn effective decision strategies through trial and error. It is used for a wide array of tasks in learning setting, among them informing hint giving \cite{fahad2023reinforcement}. For instance, previous research has explored multi-armed bandit (MAB) approaches to deliver personalized explanations \cite{williams2016axis} and to optimize the sequencing of tasks \cite{clement2014online}.

Reinforcement learning was also demonstrated to be able to support Self-regulated learning (SRL), a concept closely related to metacognitive awareness \cite{osakwe2024towards}. While not including human participants, the study could show that RL techniques like Q-learning combined with pedagogically grounded reward shaping could accurately learn strategies such as SRL. Furthermore, contemporary research could demonstrate that reinforcement learning can be used to provide adaptive pedagogical support to students by adapting difficulty and feedback, ultimately enhancing the learning effect and real-time response ability \cite{feng2024application, ruan2024reinforcement, li2024improvement}. Another approach is to leverage deep reinforcement learning (DRL) to provide metacognitive support that specifies when and how to use different learning strategies, which also significantly improves learning \cite{abdelshiheed2023leveraging}.

To utilize the advantages of GenAI in educational contexts, we thus investigate how the timing of access to GenAI can be adaptively determined via RL-based steering.  This represents a novel perspective that, to our knowledge, has not yet been systematically explored in the context of adaptive educational support. This approach enables a rigorous empirical evaluation of the impact of adaptive access timing on learning performance. 

\section{Derivation of Hypotheses}

Building on the theoretical frameworks above, we derive three hypotheses that operationalize our research questions into testable predictions.

Common theories such as metacognitive theory \cite{flavell1979metacognition}, CLT \cite{sweller1988cognitive} or the productive failure framework \cite{kapur2008productive} suggest that the structuring of external assistance in learning can significantly impact learning gains. We therefore hypothesize that modelling these theories in the RL-agents reward function and consequently have them inform GenAI use could enhance objective learning outcomes. 

\begin{center}
\fbox{\begin{minipage}{0.95\linewidth}
\textbf{H1}: Learners interacting with an RL-based policy will achieve 
significantly higher learning outcomes, measured by performance on a 
standardized post-test of the target instructional content, than those 
with naïve baseline policies (always allow / never allow).
\end{minipage}}
\end{center}

Prior results by Fernandes et al. \cite{fernandes2025ai} suggest that GenAI use can negatively impact accuracy of self assessment. Based on these findings, we hypothesize that an RL approach can enhance this accuracy by only strategically allowing GenAI use.

\begin{center}
\fbox{\begin{minipage}{0.95\linewidth}
\textbf{H2}: Learners interacting with an RL-based policy will show significantly higher metacognitive accuracy of 
their own post-test performance than those interacting with naïve baseline policies (always allow / never allow).
\end{minipage}}
\end{center}

Across past implementations of ITS, metacognitive guidance is generally explicit, requiring learners to articulate their thinking or engage in structured reflection \cite{arroyo2014multimedia, roll2011improving, azevedo2022lessons}. However, emerging research suggests that implicit scaffolds, such as delaying assistance or structuring opportunities for productive struggle, can also influence metacognitive regulation without directly prompting reflection \cite{kapur2014productive}. Motivated by this distinction, we investigate whether a minimal intervention, timing the availability of AI-based assistance, can support metacognitive engagement without relying on explicit reflective prompts.
\begin{center}
\fbox{\begin{minipage}{0.95\linewidth}
\textbf{H3:} Learners interacting with an RL-based policy will show significantly more positive change in 
self-reported metacognitive awareness of working with AI than those interacting with naïve baseline policies (always allow / never allow).
\end{minipage}}
\end{center}

\section{Methods}
\label{chap:methodology}

\subsection{Experimental Design}
\begin{figure*}[!ht]
  \includegraphics[clip,width=\textwidth]{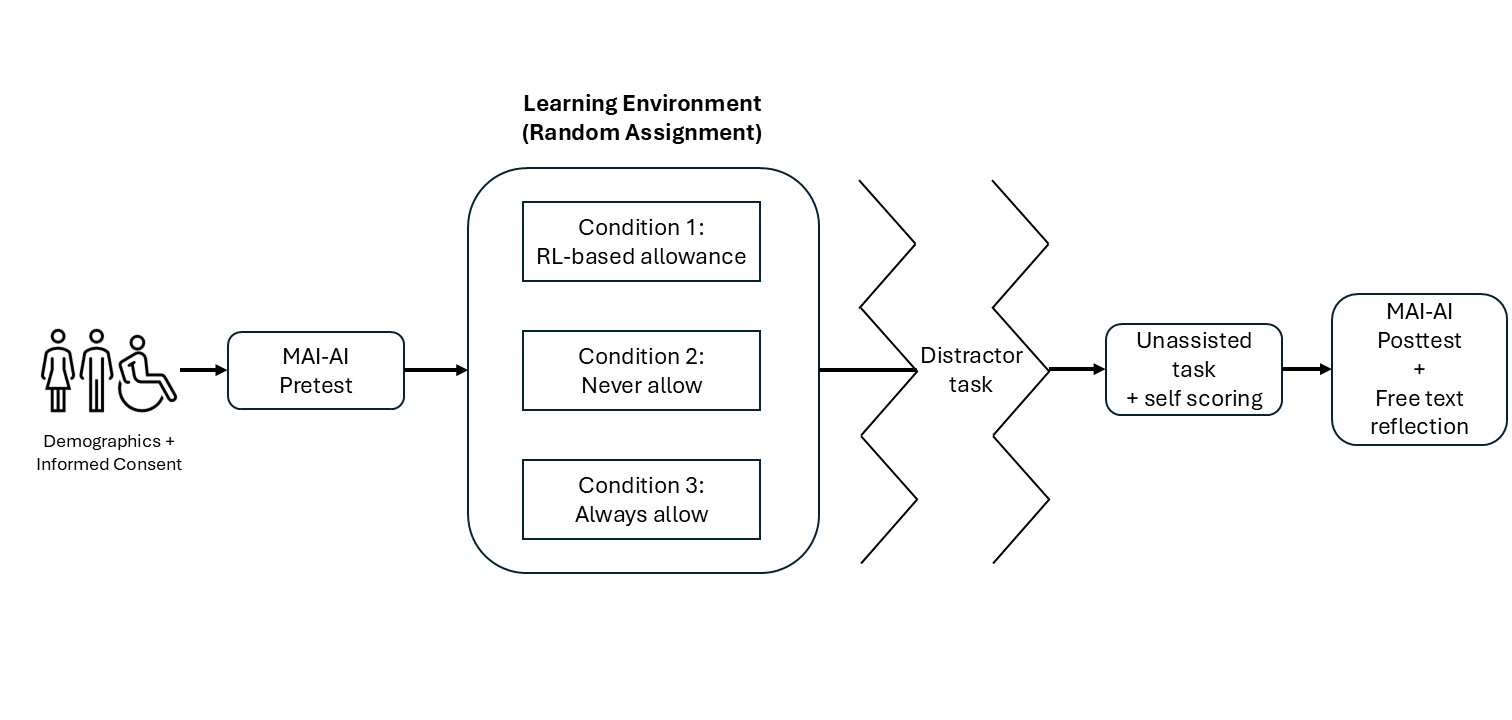}%
\caption{Visualization of Experiment Design}
\label{fig:design}
\end{figure*}

To statistically evaluate the defined hypotheses we employed a 3x1 factorial design using the policy type as an independent variable. The three conditions as illustrated in Figure \ref{fig:design} were (1) the RL agent decides at which point the learner should be allowed to use GenAI in the task, (2) the learner is never allowed to use GenAI, and (3) the learner is always allowed to use GenAI. The individual score in the task unassisted by GenAI after the unrelated activity, the alignment between this score and the students self evaluation and the pre-post test difference of the Metacognitive Awareness Inventory for Artificial Intelligence (MAI-AI) scales, a questionnaire to assess metacognitive awareness when working with AI, were treated as dependent variables. Additionally, time on each task, LLM logs and free text reflection on participant's work with the system were collected for exploratory and qualitative analysis. To avoid carry over effects, the experiment used a between-subject design. 

\subsection{Materials}

\subsubsection*{The intelligent tutoring system}

Throughout the experiment, participants interacted with a custom-built Intelligent Tutoring System (ITS) that integrates all procedural steps except the distractor task. The ITS design is inspired and significantly builds upon the system developed by Koch et al. \cite{koch2025cultural}. 
It is implemented using an AngularJS architecture and deployed as a locally hosted web interface, which includes comprehensive logging functionalities for capturing all relevant interaction data. An example screenshot of the ITS is provided in Figure \ref{fig:its} in the Appendix. The experiment was run on standardized laboratory computers to mitigate confounding effects of using personally owned computers. All devices used identical desktop computers running Windows with 23.8'' screen sizes. The ITS was accessed via Google Chrome (v143.0.7499.40), and screen resolution was fixed at 1920 × 1080 pixels.

\subsubsection*{Metacognitive awareness questionnaire (MAI-AI)}

The MAI-AI is a questionnaire to assess metacognitive awareness in relation to working with generative AI \cite{pau2026metacognitive}. It comprises 19 items on a 5 point Likert scale distributed across the two subscales \textit{knowledge of cognition} and \textit{regulation of cognition}. The instrument is derived from the Metacognitive Awareness Inventory (MAI) \cite{schraw1994assessing} and is currently under validation at Universitat Pompeu Fabra. 
The used version can be found in the additional material.

\noindent
\subsubsection*{Learning task}

A central problem many young adults in Europe face is the potentially negative influence of social media on their mental health and self-image. Studies by the European Union could show that excessive social media usage (i.e. more than three hours per day) is associated with negative mental health outcomes, specifically depression and anxiety \cite{bertoni2025social}. 

While the topic is highly relevant to students' daily lives, Chung \cite{chung2025knowing} found that young people had received very little formal instruction on how social media algorithms work during their secondary education, indicating a lack of prior knowledge. Additionally, the topic strikes a balance between conceptual complexity and accessibility, making it well suited for a text-based learning experience.

Accordingly, the learning task revolves around the issue of social media effects on self-image. To ensure its effectiveness, included content was adapted from extensively tested workshops developed by the National Health Association (NHS) in the UK \cite{duffy2022} and Canada's Centre for Digital Media literacy \cite{mediasmarts2024}. 

Overall during their work with the learning environment, participants have to complete three tasks, each consisting of a short text (approx. 650 words) on social media and self-image and three multiple choice questions concerning the text. Each multiple choice question has either two or three correct answers and after answered correctly unlocks the next multiple choice question. This is important for the RL-agent that assumes strictly linear task completion for training purposes. To facilitate learning some information is repeated between tasks.

The unassisted task after the work with the learning environment and completion of the distractor task also consists of nine similar multiple choice questions, with three each regarding the content of the previous tasks in the learning environment. However, here the order of completion does not matter. 

\noindent
\subsubsection*{AI systems}

The RL agent utilizes Proximal Policy Optimization (PPO) \cite{schulman2017proximal} to learn a policy that determines when students should be allowed to use GenAI during the learning task. PPO is a reinforcement learning algorithm designed to find an optimal policy in a sequential decision-making framework, typically formalized as a Markov Decision Process (MDP) which is commonly defined as:
\begin{equation}
M = (S, A, P, R)
\end{equation}
where:
\begin{itemize}
    \item $S$ is the set of states all possible states $s$. In our case all $s\in S$ are four-tuple that hold:
    \begin{itemize}
        \item $s_{fa}$: The failed attempts on current question
        \item $s_t$: The time on the current task 
        \item $s_{gai}$: A tuple that for each previous task holds a boolean describing whether GenAI was used or not
        \item $s_{cu}$: Current understanding defined as the number of failed attempts on the current task (aggregated version of $s_{fa}$)
    \end{itemize}
    \item $A$ is the sets of actions the agent can perform. In our case the action space is a set of two possible actions $A=\{\text{"allow AI usage"}, \text{"continue denying GenAI usage"}\}$
    \item $P : S \times A \to \Delta(S), P(s_{t+1} \mid s_t, a_t)$ can conceptually be understood as the probability that action $a$ in state $s$ at time $t$ will lead to state $s'$ at time $t + 1$. 
    \item $R: S \times A \to \mathbb{R}$ is a function that, given a state, action pair outputs the reward the agent gets. In our case it is defined as: 
    \begin{equation}
    R(s,a) = R_{\text{success}} + R_{\text{time}} + R_{\text{MT}} + R_{\text{PF}} + R_{\text{CLT}}
    \end{equation}
    The components are conceptually explained in Table \ref{tab:reward} For a complete formal definition please refer to the Appendix.
\end{itemize}

\begin{table*}[!t]
\caption{Reward components and their pedagogical grounding}
\label{tab:reward}
\centering
\begin{tabular}{lll}
\hline
\textbf{Component} & \textbf{Grounding} & \textbf{Effect} \\
\hline
$R_{\text{succ}}$ & Learning goals & Rewards task completion \\
$R_{\text{time}}$ & Efficiency & Penalizes long trajectories \\
$R_{\text{MT}}$ & Metacognition \cite{rohrer2020randomized, ostrow2015blocking} & Rewards alternating AI use \\
$R_{\text{PF}}$ & Productive failure \cite{kapur2008productive, darabi2018learning} & Rewards post-struggle access \\
$R_{\text{CLT}}$ & Cognitive load \cite{risko2016cognitive} & Penalizes premature access \\
\hline
\end{tabular}
\end{table*}

Only the reward for the chosen action is observed, which guides learning toward desirable behaviour \cite{sutton2018reinforcement}. For each of the three learning tasks, the agent decides at which point, if any, participants are granted access to GenAI. Once access is granted for a given task, it remains available until the subsequent task begins.

To train the agent, we simulate student behaviour using Bayesian Knowledge Tracing (BKT) \cite{corbett1994knowledge} with the KT-IDEM \cite{pardos2011kt} and multiple retake extension \cite{bhatt2020evaluating}. The model parameters are drawn from a truncated normal distribution centred on estimates derived from a pilot study (n=9). This allows the agent to interact with realistic student models and learn policies that generalize across diverse learner profiles. Post-hoc analysis revealed that assumptions of model parameters generalized reasonably well to initial parameter assumptions.

Through iterative PPO updates on these simulated interactions, the agent learns a policy that maximizes long-term expected learning gains. Findings reveal that in the given context PPO outperforms similar algorithms like Deep Q-Networks (DQN) \cite{mnih2013playing} and Advantage Actor-Critic (A2C) \cite{mnih2016asynchronous}. 

Illustrations of BKT parameter comparison, concrete reward parameters used, and the comparison of PPO with DQN and A2C can be found in the additional material.

The GenAI available to participants for use within the learning environment is the Mistral 3 14B 2512 LLM  \cite{mistral2025ai}, accessed via the OpenRouter API \cite{openrouter2025mistral14b}. This model developed by Mistral was chosen for several reasons. First, it demonstrates strong performance on instruction-following and text comprehension tasks relative to its size. Additionally, Mistral due to its location in Europe is subject to EU data protection regulations, facilitating GDPR compliance in the handling of participant data.

\noindent
\subsubsection*{Distractor task}

A distractor task was introduced during the retention interval to prevent students from relying on short-term memory for the following unassisted task. Specifically, participants engaged in a collaborative drawing activity inspired by and functionally similar to \cite{drake2012confronting} in which each student added to a drawing and then passed the paper to the next person. The task directs students’ focus away from the study content, ensuring that later recall reflects long-term retention rather than lingering short-term memory traces. Pilot testing confirmed that the drawing-exchange activity was engaging and attention-demanding, making it suitable as a distractor task for this purpose. The distractor task is provided in the additional material.

\subsection{Participants}

Overall, 109 voluntary university students were recruited as participants via email chains to undergraduate groups, poster advertisement and personal connections. However, some data was removed due to insufficient quality resulting in N=105 participants, as described in section \ref{sec:data_analysis}. The singular inclusion criterion was that participants were aged between 18 and 65 years to comply with ethical standards. In the resulting sample, participants were aged between 18 and 39 ($M$=20.5, $SD$=2.91). For an overview of the demographic information of participants gender and education in accordance to the newest International Standard Classification of Education 2011 (ISCED; \citeauthor{UIS2012ISCED}, \citeyear{UIS2012ISCED})  please refer to Table \ref{tab:gender} and \ref{tab:education} respectively.

\begin{table}[!ht]
\renewcommand{\arraystretch}{1.50}
\caption{Gender distribution of participants}
\label{tab:gender}
\centering
\begin{tabular}{ c  c }
\hline
\bfseries Gender & \bfseries No. of participants \\
\hline
Female & 60  \\
Male & 45\\
Diverse & 0\\
\hline
\end{tabular}
\end{table}

\begin{table}[!ht]
\renewcommand{\arraystretch}{1.50}
\caption{Highest education distribution of participants}
\label{tab:education}
\centering
\begin{tabular}{ c  c }
\hline
\bfseries Highest education & \bfseries No. of participants \\
\hline
Upper secondary education & 14\\

Bachelor or equivalent & 83\\

Master or equivalent & 8\\

Different & 0\\
\hline
\end{tabular}
\end{table}

\subsection{Procedure}
The experiment took place in 9 workshop sessions with 8 to 21 participants per sessions ($M$=12.33, $SD$=5.07), which can be considered a controlled lab environment. After arriving and providing informed consent, participants first completed the MAI-AI pretest. They were then assigned to one of the three experimental conditions (all participants in each session were assigned the same condition to keep completion time approximately equal between them). Next, they worked through three tasks, each consisting of a text followed by three multiple-choice questions. To ensure that participants actively engaged with GenAI when it was available in the RL and Always condition, they were explicitly and strongly verbally encouraged at the beginning of the study by the experimenter to use AI in any manner they considered helpful when available (e.g., for translation, copy-pasting questions, or seeking clarification). After completing these tasks, participants reached a password gate that remained locked until everyone in the session had finished the distractor task. 

Once the gate was unlocked, they proceeded to the unassisted task, which consisted of nine multiple-choice questions. For each of these questions, participants rated whether they believe they answered correctly and then completed the MAI-AI post test. Finally, they were invited to answer an optional open-ended question about their subjective experience of working with the system which concluded the experiment. 

In total, participants spent between 34.81 and 74.08 minutes in the study ($M$ = 54.39, $SD$ = 6.24). 

\subsection{Measures and Operational Definitions}
The objective learning performance of participants was assessed as the number of correctly answered multiple-choice questions in the unassisted task. After completion of data collection, an error was identified in one question which was subsequently excluded from the analysis. This exclusion did not affect the statistical validity of the study. Overall, the objective score could thus range from zero to eight. No points were deducted for false answers. For each multiple-choice question, participants assessed whether they think they answered correctly or not. Their metacognitive accuracy was calculated as the number of questions for which participants’ judgments matched their actual performance (i.e. identifying correct answers as correct and incorrect answers as incorrect), and could thus also range from zero to eight. To further differentiate between overconfidence and underconfidence, a signed metacognitive accuracy score was computed as the difference between the number of answers judged as correct and the number of answers actually answered correctly. Generally, positive values indicate overconfidence, whereas negative values indicate underconfidence. This measure was included for exploratory analyses.

To assess metacognitive awareness in the context of working with AI, the MAI-AI questionnaire was employed \cite{pau2026metacognitive}. As the instrument consisted of items rated on a 5-point Likert scale, the mean score for each of the two subscales, as well as for the total questionnaire, was used to assess metacognitive awareness. Accordingly, final scores ranged from one to five. The questionnaire was administered using a pre–post design, enabling the systematic assessment of changes in metacognitive awareness associated with the different experimental conditions. 

Additionally, behavioural indicators such as the number of AI requests, their content, and the time on individual tasks was recorded for exploratory and qualitative analysis. 

\subsection{Data Analysis Plan}
\label{sec:data_analysis}
To evaluate the proposed hypotheses, multiple statistical tests were conducted.

For H1 and H2, pairwise independent-samples t-tests were used, with experimental condition as the independent variable and individual task score (unassisted by AI) and unsigned metacognitive accuracy as the respective dependent variables. The primary contrasts of interest were RL vs. Never and RL vs. Always. 

To test H3, analysis of covariance (ANCOVA) was conducted to assess whether experimental condition influenced change in metacognitive reflection, while controlling for pre-test MAI-AI score as a covariate. Analyses were performed separately for the two MAI-AI subscales (\textit{knowledge of cognition} and \textit{regulation of cognition}).
To ensure only attentive and accurate responses were captured participants were excluded from this analysis if their MAI-AI responses met any of the following criteria: (1) straight-lining ($>$75\% identical answers), (2) extreme responding ($>$75\% of answers scored 1 or 5), or (3) low response variance ($<$0.5 SD). Overall, 14 participants were thus not considered in the analysis of H3. 

Additionally, data was excluded if a participant dropped out during the course of the experiment or failed to answer any of the questions in the unassisted task correctly, resulting in the exclusion of 4 participants and N=105.

Consistent with contemporary research standards, an $\alpha$-threshold of 0.05 was applied. As H3 was tested across multiple MAI-AI factors as well as the overall MAI-AI score, a Bonferroni correction was applied to control for the increased risk of Type I error. In contrast, no correction was applied for H1 and H2, as these only involved theory-driven comparisons between the intervention and control groups.

\subsection{Ethical Considerations}
This study received institutional ethics approval from the Universitat Pompeu Fabra ethics review body (CIREP; approval number 452). All participants provided informed consent digitally before taking part in the study and were informed about the study purpose, procedures, use of AI, and their rights as participants. Participation was voluntary, and individuals could withdraw at any time without penalty. All collected data was handled confidentially, pseudonymized at the point of collection, and stored on secure university servers with access restricted to the research team. Only anonymized data is reported or shared for scientific reuse.

\section{Results}
\label{chap:results}

\subsection{Descriptive Analysis}

The final sample consisted of N = 105 participants. Participants were assigned to the Always condition (n = 36), RL condition (n = 37), and Never condition (n = 32). Overall, there were no significant differences in the highest educational level, $\chi^2$(4) = 5.81, $p$ = .213, or age across conditions, $F$(2, 102) = 0.27, $p$ = .763. However, a $\chi^2$ test of independence showed that gender distribution significantly differed across conditions, $\chi^2$(2) = 7.82, $p$ = .020. In subsequent analyses, results controlling for gender are reported only when they differ in significance from the unadjusted analyses.

Internal consistency for the MAI-AI survey was assessed using Cronbach’s $\alpha$. For the total scale, $\alpha$ was .90, indicating excellent internal consistency. The knowledge of cognition subscale demonstrated fair reliability, with $\alpha$ = .78, while the regulation of cognition subscale demonstrated good reliability with $\alpha$ = .86. These results confirm that the MAI-AI items consistently measure a coherent underlying construct, supporting the internal validity of the instrument.

Descriptive statistics for objective post-test performance, metacognitive accuracy and self-reported MAI-AI gains are presented in Table \ref{tab:descriptive}. Means and standard deviations are shown separately for each experimental condition (RL, Never, Always), allowing additional comparison of signed accuracy, MAI-AI subscales gains, average time on task, and total AI requests across groups.

\begin{table*}[!ht]
\caption{Descriptive statistics of relevant variables}
\label{tab:descriptive}
\centering
\begin{tabular}{ c  c  c  c }
\hline
& \textbf{Always ($M$, $SD$)} & \textbf{RL ($M$, $SD$)} & \textbf{Never ($M$, $SD$)} \\
\hline
Objective post-test score (0-8) & 4.03, 1.83  & 5.11, 1.85 & \textbf{5.25}, 2.00\\

Accuracy (0-8) & 5.17, 1.36 & \textbf{5.95}, 1.51 & 5.84, 1.37\\

Signed accuracy & 1.28, 1.68 & \textbf{0.92}, 1.59 & 1.34, 1.54\\

Knowledge of cognition gain & 0.03, 0.43 & 0.09, 0.53 & \textbf{0.12}, 0.33\\

Regulation of cognition gain & -0.04, 0.48 & -0.01, 0.52 & \textbf{0.08}, 0.32\\

Combined metacognitive gain & -0.01, 0.40 & 0.03, 0.48 & \textbf{0.09}, 0.27\\

Average time on task (in min) & \textbf{2.19}, 0.57 & 2.24, 0.57 & 2.56, 0.77\\
 
Total AI requests & 11.25, 6.42 & 2.46, 3.20 & 0.00, 0.00\\
\hline
\end{tabular}
\end{table*}

Prior to analysis, Levene's test confirmed homogeneity of variances of objective performance and metacognitive accuracy across conditions, and normality was assumed via the Central Limit Theorem given group sizes exceeding 30 \cite{field2024discovering}. Additionally, homogeneity of variances and regression slopes were confirmed 
regarding the MAI-AI scores.

\subsection{Quantitative Analysis}

\begin{figure}[!ht]
  \includegraphics[clip,width=\columnwidth]{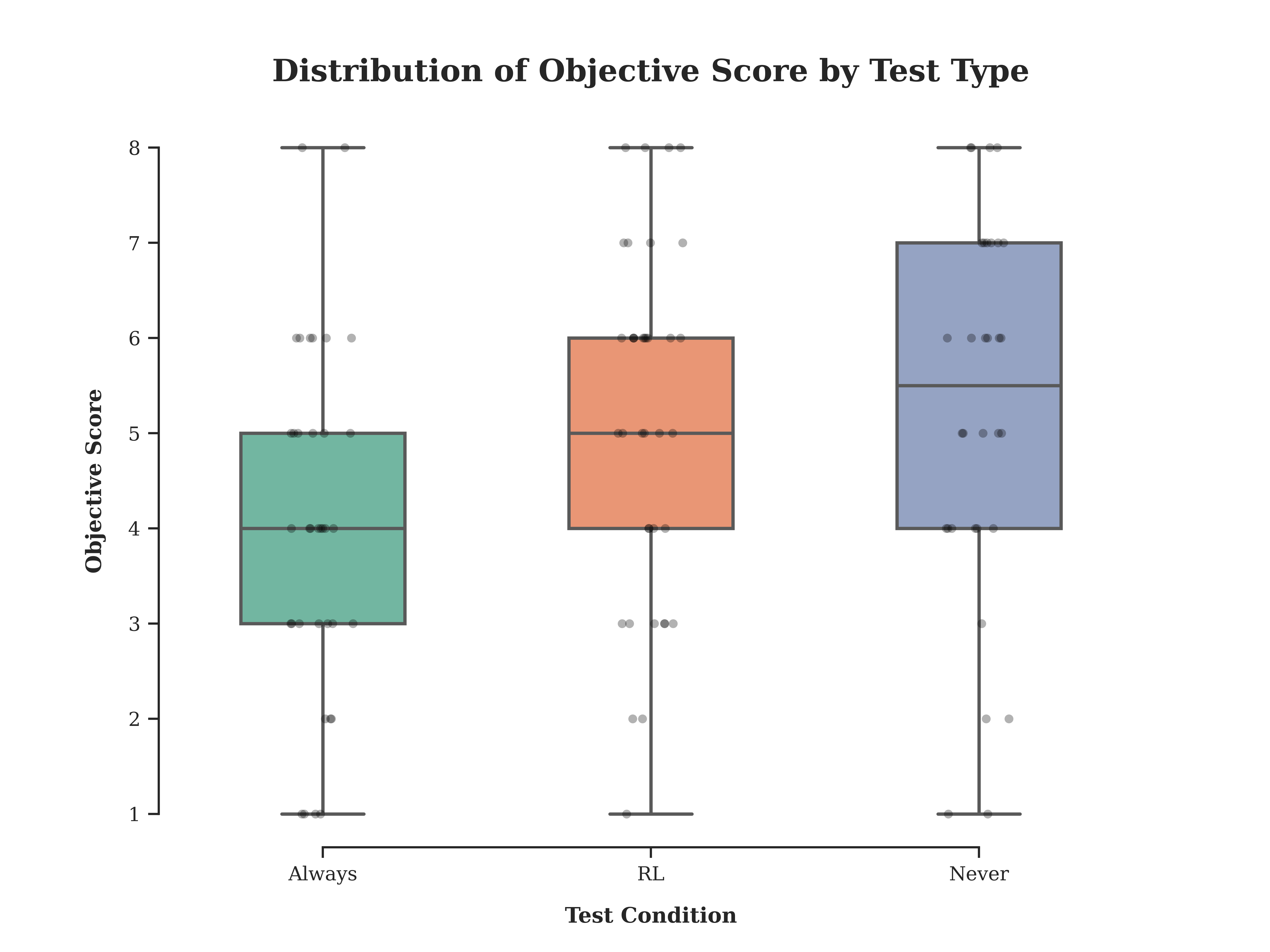}%
\caption{Boxplot comparison of the objective post-test scores}
\label{fig:scores}
\end{figure}

Results indicated that \textbf{participants in the RL condition significantly outperformed those in the Always condition regarding objective post-test scores}, $t$(71) = 2.51, $p$ = .014, $d$ = 0.59, as visualized in Figure \ref{fig:scores}. However, there was no significant difference in post-test scores between the RL and the Never condition, $t$(67) = -0.30, $p$ = .762, $d$ = -0.07. We thus partially accept H1. 

\begin{figure}[!ht]
  \includegraphics[clip,width=\columnwidth]{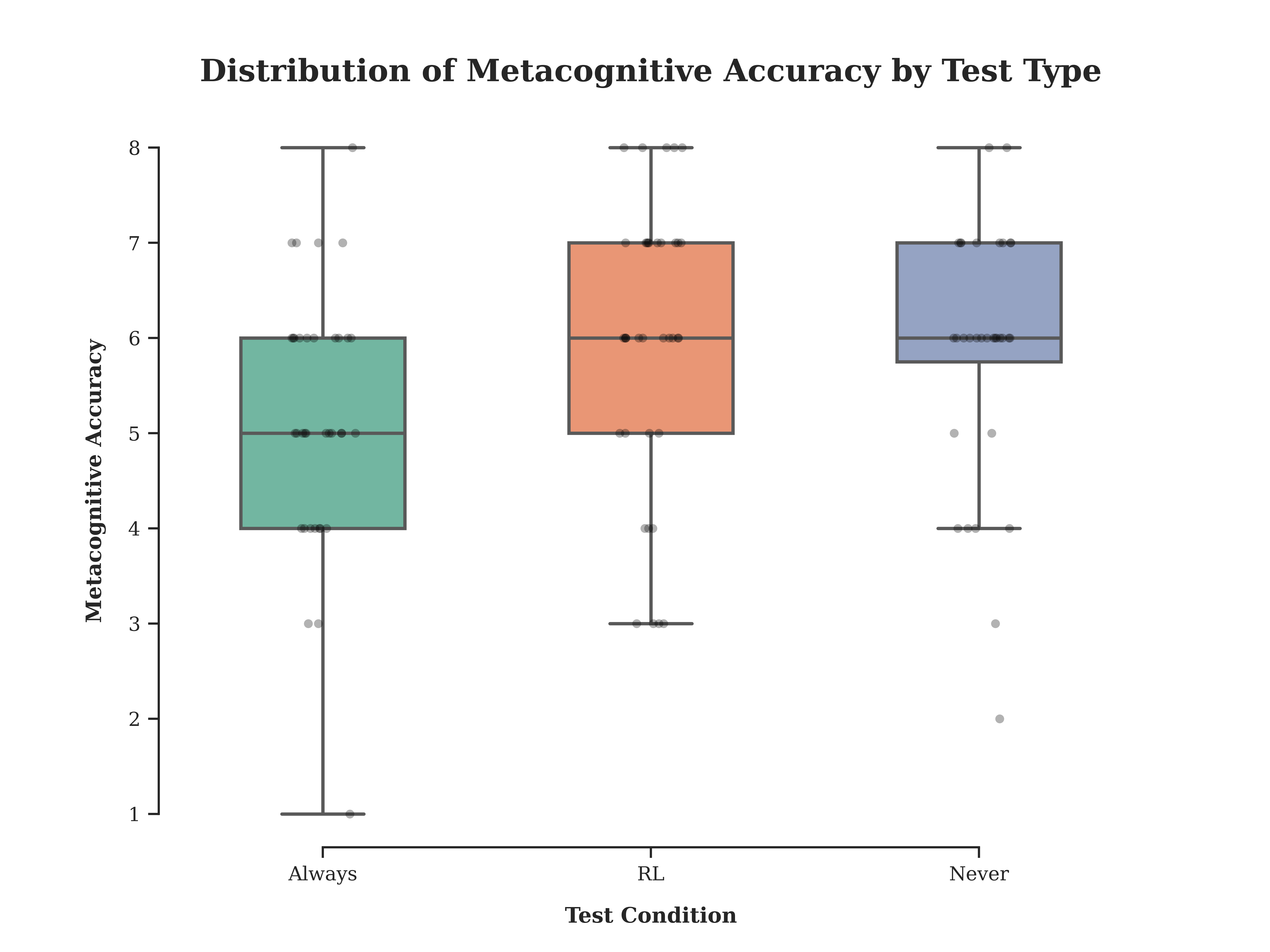}%
\caption{Boxplot comparison of the metacognitive accuracy}
\label{fig:accuracy}
\end{figure}

Additionally as shown in Figure \ref{fig:accuracy}, the data reveals that \textbf{participants in the RL condition exhibit significantly higher metacognitive accuracy than those in the Always condition}, $t$(71) = 2.32, $p$ = .023, $d$ = 0.54. On the other hand, there was no significant difference in metacognitive accuracy between the RL and the Never condition, $t$(67) = 0.29, $p$ = .769, $d$ = 0.07. Therefore, we partially accept H2. 

Separate one-way ANCOVAs were conducted to examine the effect of intervention condition on MAI-AI factors and overall metacognitive gain, controlling for pre-test MAI-AI scores. After adjusting for pre-test scores, the effect of condition was not significant for knowledge of cognition gain, $F$(2, 87) = 0.39, $p$ = .678, $\eta_{p}^2$ = 0.01, regulation of cognition gain, $F$(2, 87) = 0.86, $p$ = .427, $\eta_{p}^2$ = 0.02, and overall metacognitive gains, $F$(2, 87) = 0.86, $p$ = .425, $\eta_{p}^2$ = 0.02. We thus reject H3. 

In addition to the analysis of the hypothesis, an exploratory analysis was conducted to examine differences in both the average completion time per question and the average number of mistakes per question. Results indicated that there was no significant difference between the RL and Always conditions in terms of the average number of mistakes per question, $t$(71) = 0.68, $p$ = .498, $d$ = 0.16. However, \textbf{participants in the RL condition made significantly fewer mistakes than those in the Never condition}, $t$(67) = 2.05, $p$ = .047, $d$ = 0.51, potentially reducing frustration. This effect is not statistically significant anymore after controlling for gender, $b$ = 0.46, $p$ = .513.
Furthermore, \textbf{participants in the RL condition took marginally significantly less time per question than those in the Never condition}, $t$(67) = 1.94, $p$ = .058, $d$ = 0.47. After controlling for gender, this difference was no longer significant, $b$ = 2.74, $p$ = .769. 
No significant difference was observed in average time per question between the RL and Always conditions, $t$(71) = 0.37, $p$ = .714, $d$ = 0.09.

\subsection{Qualitative Analysis}
Beyond the quantitative analysis, we examined the qualitative material gathered through inductive thematic analysis \cite{clarke2017thematic}. This process involved reviewing the text participants submitted as prompts within the chatbot, as well as their answers to the concluding open-ended question about their experience with the system. The analysis followed multiple iterative stages and was carried out by the first author by hand. Initially, both the prompts and the open-ended responses were systematically coded, identifying recurring terms, phrases, and underlying ideas. This process relied on inductive codes (emerging directly from the data, e.g., "Laziness/Dependence on AI"). Related codes were subsequently grouped into broader, overarching organizing themes which in turn were grouped into global themes. To support this process, a thematic map was created to visualize and clarify the connections among themes and to guide their refinement. The resulting thematic structure and the full codebook are provided in the additional material. 
Finally, these themes were integrated into a cohesive narrative that brings together the key insights from the analysis. During the whole process, special attention was paid to how the distribution of themes differed between conditions. 

\noindent
\subsubsection*{GenAI Usage Between Conditions}

\begin{figure}[!ht]
  \centering
  \includegraphics[clip,width=\columnwidth]{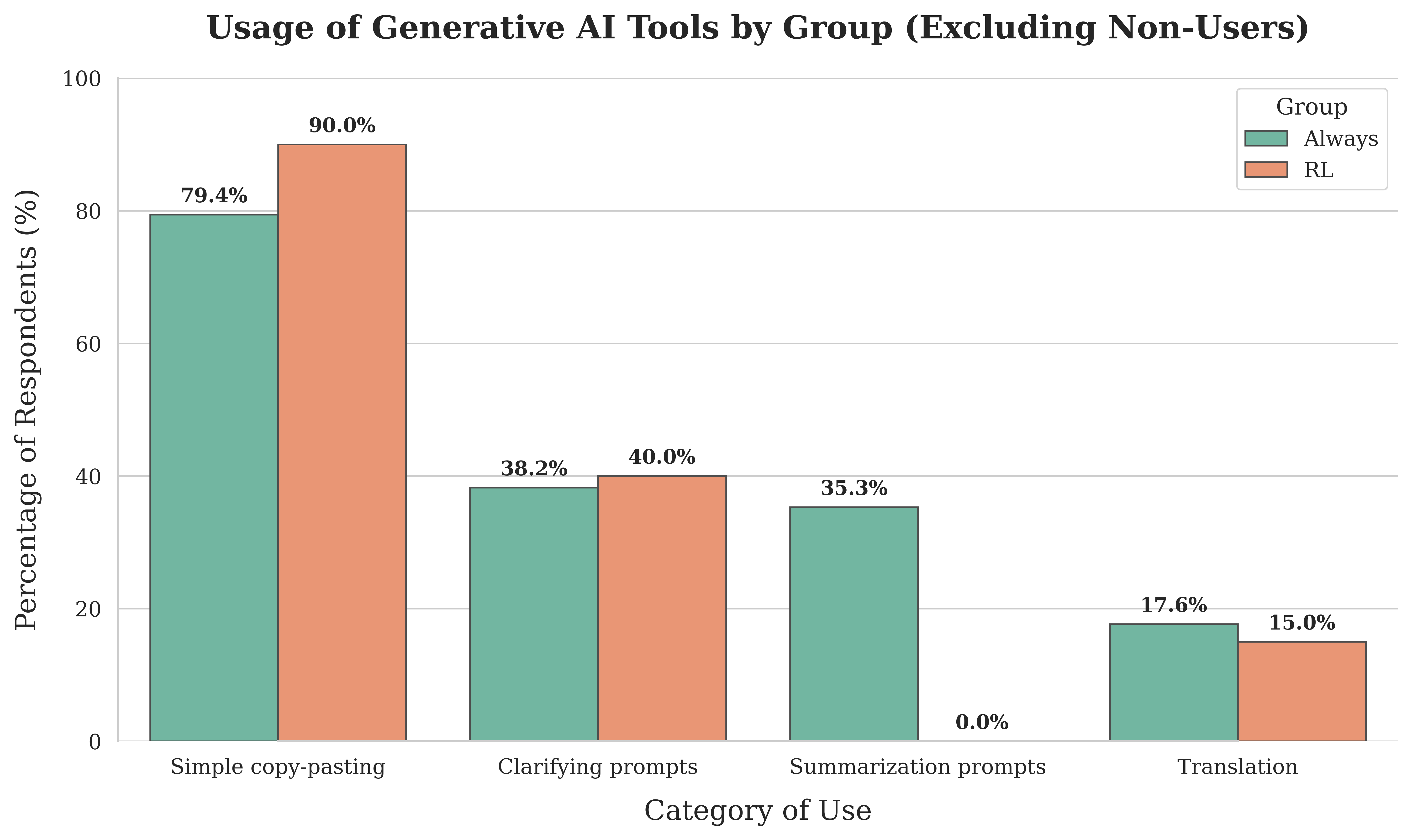}
  \caption{Comparison of GenAI usage patterns across conditions. Note that bars may not sum to 100\% as participants could use multiple strategies.}
  \label{fig:genaiusage}
\end{figure}

Overall, the usage distribution of GenAI differed strongly between the RL and Always condition. First of all, 43.2\% in the RL condition did never use AI either because the agent did not think it necessary to allow usage or because they deliberately choose not to. This is strongly opposed to the Always condition, where all participants used GenAI at least once. The remaining distribution of usage patterns excluding participants who never used GenAI is visualized in Figure \ref{fig:genaiusage}.

While both conditions exhibit a similar likelihood to use clarifying prompts or translation approaches, participants in the RL condition seem more likely to simply copy-paste questions. However, they did so more selectively than participants in the Always condition and rarely copy-pasted more than one multiple-choice question by task. Another key point is that participants in the Always condition used summarization prompts while users in the RL condition never relied on it. This is most likely due to participants in the RL condition being forced to read the text before eventually gaining access to GenAI thus already knowing and engaging with its content rendering summarizing prompts useless. 

\subsubsection*{Sentiment Towards the Own Condition}

\begin{figure}[!ht]
  \centering
  \includegraphics[clip,width=\columnwidth]{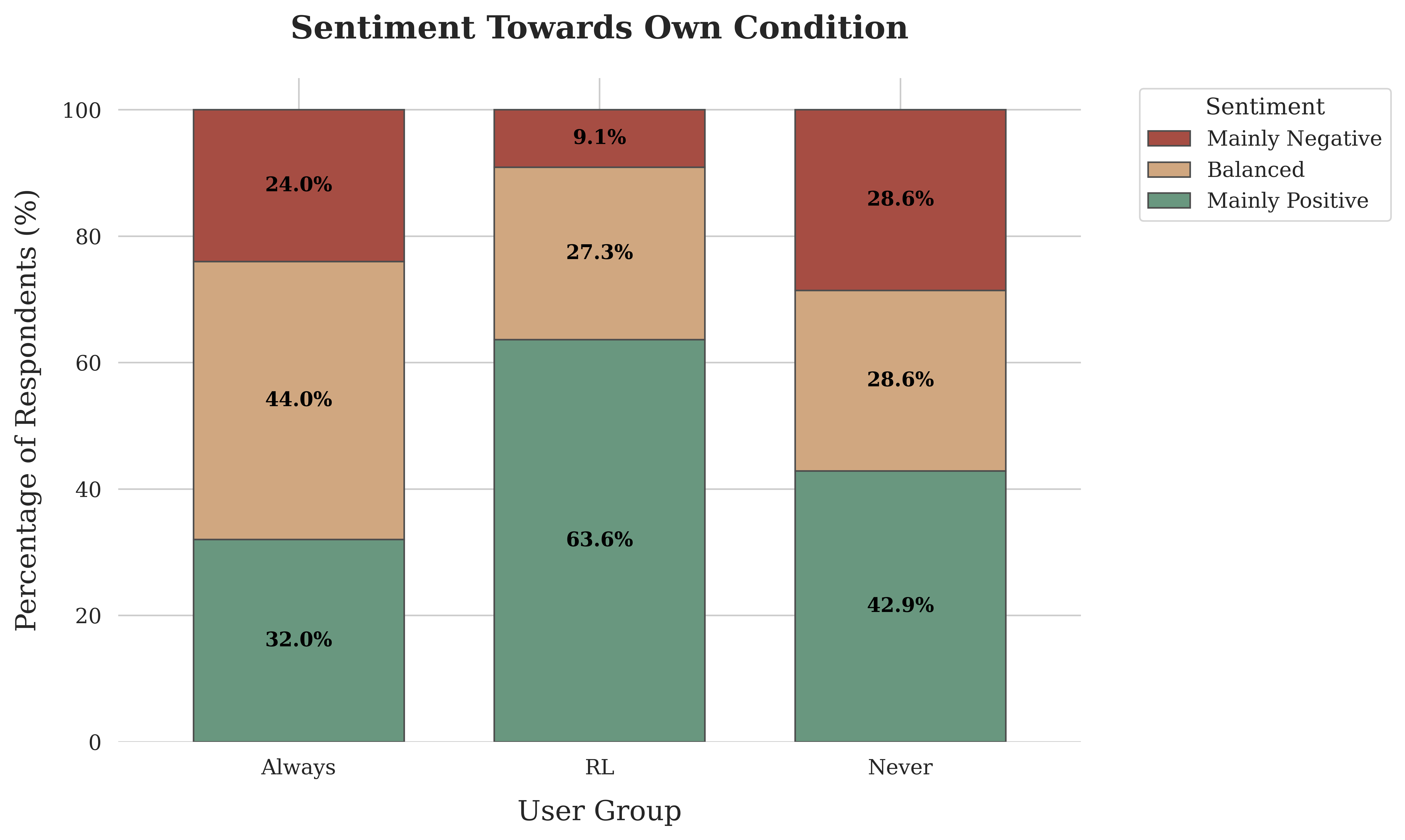}
  \caption{Comparison of sentiment towards the own condition between groups.}
  \label{fig:sentiment}
\end{figure}

Regarding the self-reflection prompts which asked about the perceived usefulness of the GenAI used in the corresponding condition, participants both reported benefits and detrimental effects of using GenAI in accordance to their condition. The distribution of the sentiment of the responses is visualized in Figure \ref{fig:sentiment}. Annotation was done by two independent reviewers reaching substantial agreement (Cohen's $\kappa$ = .75). Conflicts were resolved through discussions.

\noindent
\subsubsection*{Perceived Benefits of Using GenAI}

Many participants in the RL condition noted that \textit{"it was helpful to have [access to] the AI assistant after 'failing' [...] a few times, to not get frustrated and lose track" (RL condition)}. This \textbf{importance of initial failure} was consistently highlighted throughout this condition as one other participant put it as \textit{"It was [...] super useful to just have the AI assistance after failing because that is the whole point of learning, we need to fail to be able to learn and this enhances thinking" (RL condition)}. Another crucial point that was perceived as beneficial in the RL condition was the \textbf{motivation to try tasks by oneself first}. One participant noted that \textit{"the positive aspect of having AI assistance only sometimes, is that you have to [...] read the text [...] first, understand it and then, if you have the assistance of AI, you can use it to support what you already learnt" (RL condition)}. Another participant noted that \textit{"not having AI at all times forced me to try to work out the answers [...] before searching any type of aid" (RL condition)} but then also having access to GenAI \textit{"when I was stuck on a question, so I [...] would not get frustrated" (RL condition)}. The potential benefit of this technique was even noted in other conditions as one participant in the Always condition stated that \textit{"AI need[s] to learn how to help the user [and] notify them that he/she needs the artificial intelligence too much" (Always condition)}. Overall, some users also indicated a sustainable long-term effect of using GenAI in this way and that \textit{"at the end of the experiment [their] opinion on AI has changes and in the future [they] will try to use it in a more responsible way" (RL condition)}.

Beyond the perceived benefits in the RL condition participants noted a \textbf{general desire to use AI} independent of condition: \textit{"I believe [AI] improved my learning routine a lot" (Always condition)}. However, this desire was described as contextually dependent and one user put it as: \textit{"Maybe I would have used generative AI if this was another setting, like if I was alone in my room, and this questionnaire had a pressing delivery date" (Never condition)}.

A primary driver for this desire across conditions was the \textbf{ability to save time}: \textit{"I think the work [with the] system is interesting, since being able to use [such] tools [...] saves you a lot of time." (RL condition)}. Along the same line participants also reported that \textit{"[AI] has been really useful to not get blocked" (Always condition)}. However, this was not always seen as a positive aspect and some users did not think that in the given task \textit{"generative AI would be useful in order to learn [about] the content. It would just make the process of answering questions faster" (Never condition)}.

\noindent
\subsubsection*{Perceived Pitfalls of Using GenAI}

Concerning the detrimental effects of using GenAI during the experiment many users cited \textbf{laziness and dependence on the system} as a primary concern. Particularly in the Always condition participants noted that working with GenAI has made them \textit{"more dependent" (Always condition)} and \textit{"lazier than before” (Always condition)}. Another user noted that \textit{"I use the AI to finish fast and I know it, so if I want to learn I have to force myself to not use it" (Always condition)}.

Another key point noted in the RL and Always condition was that \textbf{perceived usefulness degraded over time} in some instances. For example, one participant stressed that \textit{"in the beginning it helped me but over time, when I was getting tired and I felt pressured to finish fast [...], it hindered my learning since I started believing AI without checking if [its answer] was stated in the text" (Always condition)}.

Overall, participants in the RL and Always conditions highlighted the subjective \textbf{feeling of diminished learning} when using GenAI. Users consistently noted, that \textit{"I think that AI hindered my learning because later I wasn't sure about any question" (Always condition)} and that \textit{"[GenAI] just gives a simple answer, not deep understanding. So I have to use it carefully" (RL condition)}. Especially \textit{"the possibility [to] just ask for the right answer" (RL condition)} was seen as a primary reason for this. 

Finally, it is also noteworthy that particularly in the RL condition the danger of \textbf{easiness to game the system} was noted by some participants, as for instance one of them indicated that \textit{"once I knew [GenAI] will [be available] after some wrong answers. I was tempted to answer wrong purposely to get [access to the GenAI] especially when I didn´t want to do the thinking on my own" (RL condition)}.

\section{Discussion}
\label{chap:discussion}

The present study investigated whether pedagogically grounded timing of access to GenAI, implemented via an RL-based agent trained on reward signals informed by metacognitive theory, cognitive offloading, and productive failure, can meaningfully outperform naive baselines in supporting learning outcomes and metacognitive development. A consistent pattern emerged across primary and exploratory analyses: strategically timed GenAI access improved objective post-test performance and metacognitive accuracy compared to unrestricted access, while exploratory comparisons further suggested reduced task errors and time on task relative to complete withholding, all without requiring explicit metacognitive prompting or structured scaffolding. Qualitative findings reinforce this picture, with participants in the RL condition responding most positively to their learning experience and spontaneously articulating the value of initial struggle. The one exception to this pattern was self-reported metacognitive awareness, where no significant between-condition differences emerged. Taken together, these results position the timing of GenAI access as a meaningful and theoretically grounded pedagogical method that preserves the cognitive benefits of productive struggle while mitigating the risks of unrestricted AI reliance to ultimately outperform unrestricted access and potentially improve over fully restricted access.

The results show clear signs that grounding the decision of when to use AI in pedagogical principles can significantly enhance objective learning gains, over having unrestricted access. This illustrates that, beyond external self-evaluation prompts, planning support \cite{azevedo2022lessons} and metacognitive prompts \cite{singh2025enhancing, xu2025enhancing} simply timing the access to GenAI can already enhance learning. Importantly this also works on unscaffolded naive tools like for instance off-the-shelf LLMs like Mistral or Claude. According to qualitative findings this improvement may be due to the ability to ask clarifying questions, understand answers deeply and translate content providing targeted individualized support. These results further align with recent findings that independent work before LLM use can be beneficial for critical thinking given sufficient time on the task \cite{zhi2026investigating}.

Furthermore, the same intervention that improved learning also recalibrated metacognitive accuracy, suggesting these aren't independent effects. Past research by Fernandes et al. \cite{fernandes2025ai} highlighted that users are unable to accurately assess their performance when using GenAI when compared to not relying on it. While reproducing this finding, our results strongly indicate that only strategically allowing the use of GenAI recalibrates cognition and performance leading to similar outcomes when compared to never using GenAI. According to the productive failure framework \cite{kapur2008productive}, this alignment may be due to understanding a task thoroughly on ones own and failing before receiving structured support, leading to an increased understanding of ones own skills.

While strategically allowing GenAI access could thus clearly improve over unrestricted access, it also had significant benefits over complete withholding in exploratory analysis. For instance, participants in the RL condition made significantly fewer mistakes than the ones with fully restricted access. Qualitative findings further indicate that this reduced frustration in learners. Through the lens of cognitive load and productive failure this finding aligns with Kapur's observation that productive struggle enhances learning only insofar as cognitive load and frustration remains manageable and does not cause learners to disengage entirely \cite{kapur2014productive}.

Additionally, participants in the RL intervention required marginally less time per task than learners without access to GenAI. While this result should be interpreted with caution, it is in line with findings by Noy \& Zhang \cite{noy2023experimental}. At first glance, such acceleration may appear beneficial. However, some research suggests that it may bypass important cognitive processes that typically occur during slower, more deliberate learning by offloading them to GenAI \cite{singh2025protecting}. However, given that the objective post-test scores were approximately equal between the RL and Never conditions, this interpretation seems unlikely in the present case and rather a reflection of higher learning efficiency.

Qualitative analysis further revealed that selective access to GenAI was viewed by participants as most positive. Participants in the RL condition valued the importance of initial failure and the motivation to try tasks by oneself first. This aligns well with the used pedagogical principles of productive failure and cognitive load theory. Furthermore, learners with unrestricted access commonly talked about feelings of laziness and dependence on the GenAI system, supporting the notion of metacognitive laziness \cite{fan2025beware} and external reliance \cite{hou2025role, zhai2024effects} commonly discussed in literature.

Overall, the results thus indicate that strategic timing of GenAI access via the RL condition outperforms unrestricted access, while also potentially improving over fully restricted access. This points to the importance of integrating pedagogical insights with human-AI interaction in education \cite{mittal2024comprehensive}. Especially, when combined with other scaffolding mechanisms such as metacognitive prompting or adaptive feedback, strategic access timing could therefore lead to deeper and more durable learning gains. However, it should be noted that as common for scaffolding techniques surrounding reflective GenAI usage final learning outcomes also partly depend on the quality of the underlying LLM. If the model provides low-quality or misleading responses when access is granted, the pedagogical benefit of strategic timing may be weakened.

Results showed no between-group differences regarding metacognitive knowledge and regulation as assessed by the pre-post test of the MAI-AI. There could be a number of reasons for this finding. First, using self-assessment tests for metacognitive measurements is widely criticized in literature as metacognitive knowledge and regulation is often not adequately related to
metacognitive behaviour \cite{craig2020evaluating}. Additionally, research suggests that we may hold inaccurate beliefs about our metacognitive monitoring capacity \cite{double2025survey}, further reducing the effectiveness of self-assessment measures. These results might thus represent more of a design insight, informing future work, than a limitation. Another reason could be that students simply did not engage more in metacognitive reflection in the RL condition than in the Never and Always condition beyond a recalibration of cognition and performance. This raises the question of to what degree the findings can be interpreted through the lens of metacognition. A final option concerns the fact that even though the MAI-AI shows good internal reliability and was used in previous studies \cite{pau2026metacognitive} it still awaits validation. While the high internal consistency indicates that the items measure a coherent construct, it does not guarantee that the instrument accurately captures the intended dimensions of metacognitive knowledge and regulation. It is therefore possible that the null results of H3 partly reflect limitations of the instrument rather than a true absence of between-condition differences.

\subsection{Implications}

The finding that strategically timed access outperforms unrestricted access, with exploratory evidence that it may also improve on fully restricted access, offers strong empirical support for access timing as a novel and effective mechanism to scaffold learning with minimal invasiveness. Taken together with the recalibration of metacognitive accuracy in the RL condition, this suggests that some dimensions of metacognitive benefit may emerge without explicit and structured prompting, as commonly assumed in ITS research. Additionally, the results offer support for the intersection of cognitive offloading and productive failure. They also suggest that the core ideas of the productive failure framework which was originally developed for a non-AI context retains its expressiveness when applied to GenAI timing decisions, extending its theoretical scope. Finally, the null-results on changes in metacognitive engagement together with the increase in metacognitive accuracy raise a theoretically important question about the relationship between metacognitive accuracy and metacognitive awareness as potentially distinct constructs.

For educators and instructional designers the results suggest that simply adapting the timing of access to GenAI, without requiring additional infrastructure or explicit pedagogical prompting, can meaningfully improve learning outcomes and self-assessment accuracy. This is a low-cost easily traceable intervention that is compatible with off-the-shelf tools like Claude or Gemini, that can be readily deployed in classrooms without requiring custom-built AI systems. For policymakers navigating the educational landscape the results advice against full restrictions or unrestricted access as viable defaults but rather advocate for a balanced approach that allows educators to strategically use GenAI when pedagogically desirable. For developers these results empirically demonstrate that a RL agent trained on a pedagogically grounded reward function with simulated student data can learn meaningful access policies, thus offering a traceable framework to integrate similar mechanisms into existing platforms.

\subsection{Limitations}

Nevertheless, this study suffers from a number of limitations. First, the student modelling to simulate data was based on limited interactions (n=9) and post-hoc reestimation revealed moderate deviations for some parameters and skills. However, these selective differences had minimal effect on the learned policy as the dominant role of pedagogical reward shaping and the use of truncated normal sampling during training buffered the policy against moderate parameter misspecification, as evidenced by the significant between-group effects observed. Second, the majority of participants spoke a Romance language other than English as their first language, while the experiment itself was conducted in English, potentially leading to a language bias. Third, this study only investigates the temporary use of the system and does not longitudinally track learning over a prolonged period of time. This is particularly noteworthy as a closer familiarity with the system might lead to learners being able to game the system and adapt their behaviour to gain faster access to GenAI. Fourth, qualitative data analysis is inherently subject to interpreter bias. Even though measures were taken through the development of a thematic map, extensive code book, and measurements of inter-rater reliability the possibility for self-serving and confirmation bias, particularly regarding the sentiment of participants towards their own condition, remains. Finally, the learning tasks in this study concerned one specific topic (social media and mental health) and text comprehension, potentially limiting how far findings can be extended to other content areas or skill types.

\subsection{Future Work}

Building on the findings and limitations of this study, future work should employ ablation studies to investigate which components of the pedagogically grounded reward function and thus pedagogical theories were primarily responsible for the observed effects. Additionally, given the established importance of intrinsic motivation in GenAI-mediated learning, subsequent research could move beyond metacognitive awareness as the primary outcome measure and explicitly monitor motivational dynamics across different access timing conditions. Furthermore, replicating and extending these findings beyond controlled laboratory environments for instance, in classroom settings over longer time horizons would substantially strengthen the ecological validity and practical transferability of the results. Finally investigating how teachers can facilitate these findings and co-develop appropriate tools could further increase the impact of findings while keeping humans in the loop. 

\section{Conclusion}
The present study shows that pedagogically grounded timing of access to GenAI, implemented via an RL-based agent utilizing reward shaping derived from metacognitive theory, cognitive load theory, and productive failure, can meaningfully outperform unrestricted access by simply timing GenAI access during learning. Furthermore, exploratory analysis indicates an improvement over fully restricted access. We show how the system enhances objective performance and metacognitive accuracy compared to unrestricted access, while also potentially reducing errors and time on task compared to complete withholding access. These findings imply that educators do not necessarily need to rely on complex scaffolding techniques but that a pedagogically grounded decision of when to allow access already represents a low-cost, scalable solution that facilitates learning. This opens up a new research area for how this method can be facilitated by educators to improve long-term learning and practically implemented in intelligent tutoring systems. Ultimately, these results suggest that the challenge of integrating GenAI into education may not lie in restricting it, but in learning when to allow it.

\section*{Acknowledgments}
\noindent
Special thanks are attributed to Batuhan Sayis 
for his assistance in receiving the ethics approval.

\subsection*{Funding}
This work was supported by the Spanish Research Agency (AEI) under PID2023-146692OB-C33 and 2023-26 CEX2021-001195-M by MICIU/AEI/10.13039/501100011033. DHL (Serra Húnter) acknowledges support by ICREA Academia.

\subsection*{Additional Material}

Additional material can be found on OSF under \\\url{https://osf.io/ezs3b/overview?view_only=c3f862c43eb641aa83053c47e69c60ea}.

Additionally, the code for the tutoring system, can be found on GitHub under \\
\url{https://github.com/Janne-ro/learning-when-to-help.git}.

\subsection*{Generative AI Use}
This paper was developed with support from generative AI (ChatGPT-5.1 and Claude Sonnet 4.6) for grammar and stylistic polishing during document refinement. While the authors initially wrote all of it, directed the whole process and made all critical revisions, the AI’s stylistic input is woven throughout. All changes made by the AI were minor and reviewed by the human author. The authors take full responsibility for the content of this manuscript.


\printbibliography


\appendix

\clearpage
\section*{Appendix - Pedagogical reward shaping}
\label{app:reward}
The reward function is a crucial part of a RL agent that defines which actions are desirable for the RL agent to take to maximize long term discounted reward. In line with previous research which employed pedagogical reward shaping we base the behaviour of the RL agent in educational principles \cite{barnes2007toward, osakwe2024towards}. Specifically, we consider metacognitive theory, CLT, and productive failure.

We construct the instantaneous reward at timestep $t$ which is included in the state $s$ as the sum of a small set of interpretable components:
\begin{multline}
R(s,a) = R_{\text{success}}(s) + R_{\text{time}}(s) + R_{\text{MT}}(s,a) + R_{\text{PF}}(s,a)\\ + R_{\text{CLT}}(s,a)
\label{eq:total_reward}
\end{multline}

\subsection*{Task success and temporal efficiency}
A high positive reward is given when the student successfully completed the current task (to encourage achieving learning goals) in the 5 second time interval after the last action, while a small negative timestep penalty encourages efficiency and discourages trivially long trajectories:
\begin{align}
R_{\text{success}}(s) &= c_{succ} \cdot \mathbf{1}_{{\text{task completed at } t}}\\
R_{\text{time}}(s) &= -c_{time},
\end{align}
with $c_{succ} \in \mathbb{R}^{+}$ large relative to per-step penalties $c_{time} \in \mathbb{R}^{+}$. These terms implement the standard objective of maximizing long-term discounted task attainment while preferring shorter, more focused interactions.

\subsection*{Metacognitive Theory}
Metacognitive interventions often rely on alternating experience and reflection (for example, exposing learners to contrasting approaches) \cite{rohrer2020randomized, ostrow2015blocking}. To encourage such reflective sequencing, we provide a small incentive for actions that present a contrast relative to the student’s history.
\begin{equation}
R_{MT}(s, \text{"allow AI usage"}) = 
\begin{cases}
c_{MT} \text{, if } s_{gai} \cap \text{True} = \emptyset\\
0 \text{, else}
\end{cases}
\end{equation}
Conversely: 
\begin{equation}
R_{MT}(s, \text{"don't allow AI usage"}) = 
\begin{cases}
c_{MT} \text{, if }  s_{gai} \cap \text{False} = \emptyset\\
0 \text{, else}
\end{cases}
\end{equation}
with $c_{MT} \in \mathbb{R}^{+}$

\subsection*{Productive Failure (PF)}
\label{sec:rl-pf}
Productive failure posits that an optimal amount of struggle can benefit later learning \cite{kapur2008productive}, but excessive failure becomes detrimental \cite{darabi2018learning}. We therefore shape rewards to (1) discourage giving additional help too early, (2) promote offering help after engaging in productive struggle, and (3) strongly penalize prolonged or excessive failure without assistance.
\begin{equation}
R_{\text{PF}}(s,\text{"allow AI usage"}) = 
\begin{cases}
-c_{PF} \text{, if } s_{fa} = 0\\
\alpha \cdot s_t \text{, if } 1<=s_{fa}<=2\\
\beta \cdot s_t \text{, if } s_{fa} > 2
\end{cases}
\end{equation}
with $\alpha,\beta, c_{PF} \in \mathbb{R^{+}}$. We additionally define $\forall s:R_{PF}(s, \text{"don't allow AI usage"}) = 0$. This suffices, as in a binary action space the difference in immediate reward between the two action is the relevant factor for action selection \cite{sutton2018reinforcement}.

\subsection*{Cognitive Load Theory}
CLT warns against premature or excessive cognitive offloading (e.g. relying too early on external aids) especially during the early phases of learning \cite{risko2016cognitive}. We therefore penalize actions that offload cognition (here allowing generative AI assistance) when the learner is still in an early stage of the task.
\begin{equation}
R_{CLT}(s,\text{"allow AI usage"}) =  
\begin{cases}
-\delta \cdot (T_{s_t} - s_t) \text{, if } s_t<T_{s_t}\\
0 \text{, else}
\end{cases}
\end{equation}
where $T_{s_t}, \delta \in \mathbb{R}^{+}$ and $T_{s_t}$ denoting the threshold in seconds for being in an early stage of learning. Following the same logic as in the previous subsection, we define $\forall s: R_{CLT}(s, \text{"don't allow AI usage"}) = 0$.

\subsection*{Practical notes on parameterization and training}
To ensure that the shaped rewards produce the intended instructional behaviour, multiple offline simulations were run. Additionally, a second pilot study with n = 7 participants was conducted to further fine tune the parameters. The concrete values of reward- and hyperparameters can be found in the additional material.

\begin{figure}[!h] \includegraphics[clip,width=0.95\columnwidth]{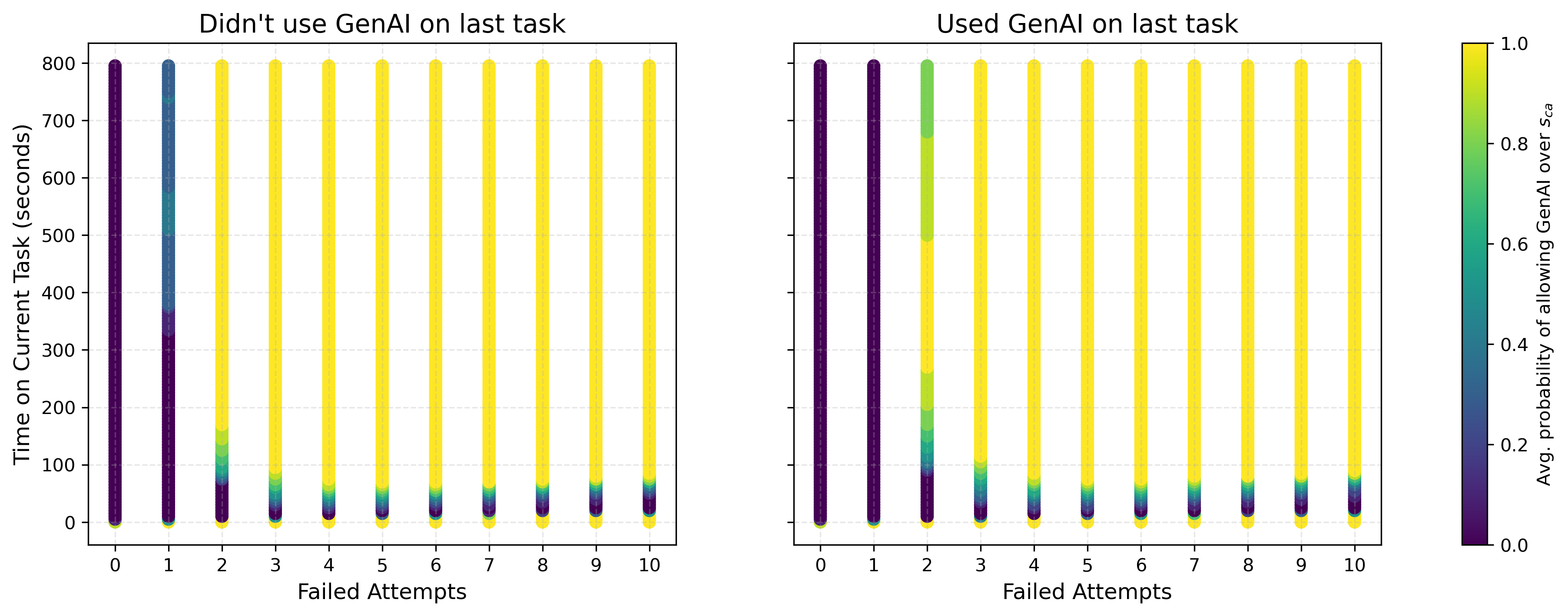}
\caption{Visualization of the final policy in the case of the agent being in the second task}
\label{fig:policy_comparison}
\end{figure}

The finally derived policy is visualized in Figure \ref{fig:policy_comparison}.

\newpage

\section*{Screenshot of the ITS}
\begin{figure}[!ht]
  \centering
  \includegraphics[clip,width=\columnwidth]{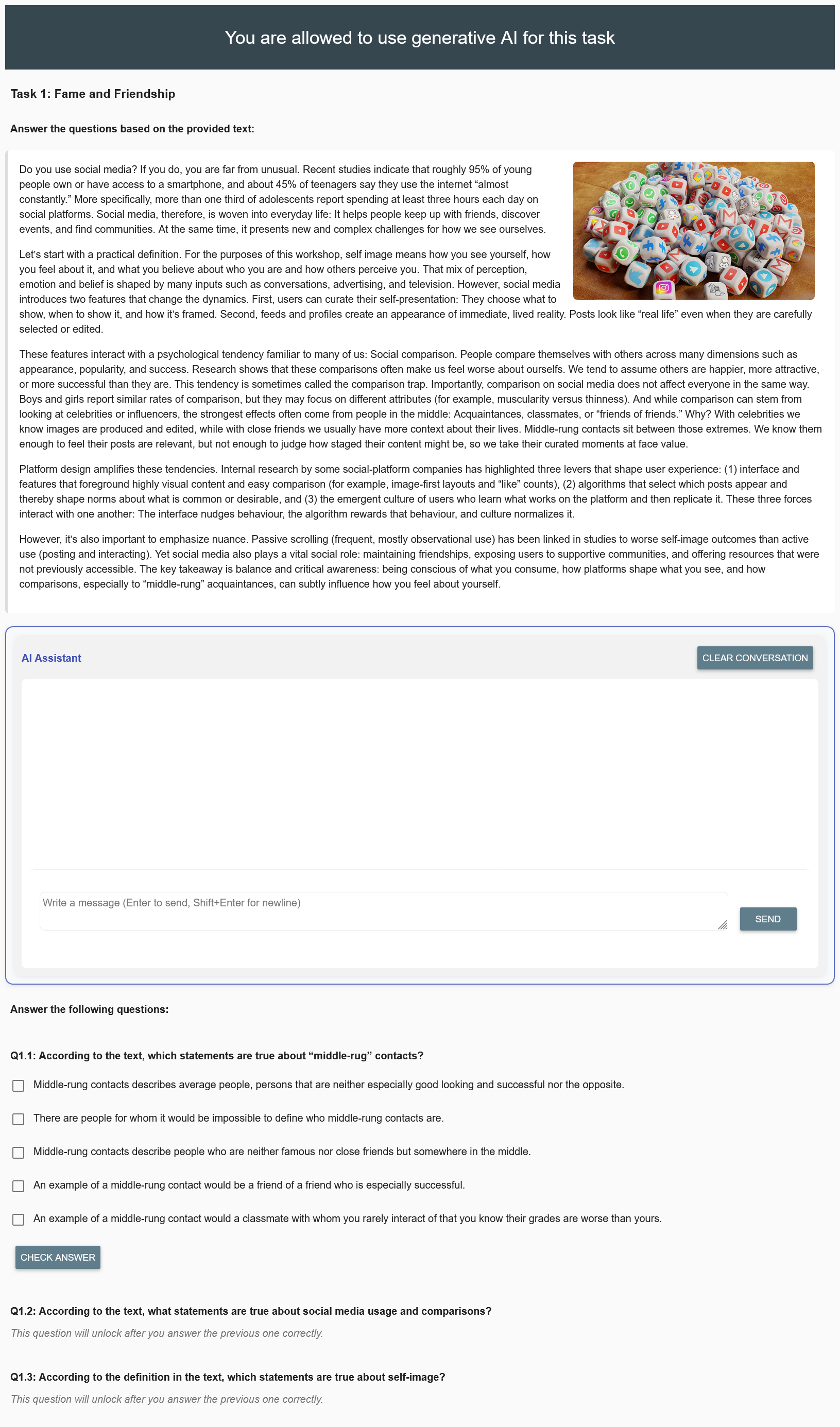}
  \caption{Example screenshot of the ITS}
  \label{fig:its}
\end{figure}

\end{document}